\documentclass[twocolumn,floatfix,superscriptaddress,amsmath,showpacs,aps,pre]{revtex4}
\begin{document}
\title{Functional integral approach for multiplicative stochastic processes}
\author{Zochil Gonz\'alez Arenas}
\affiliation{Instituto de Cibern\'etica, Matem\'atica y F\'\i sica (ICIMAF)\\
Calle 15 \# 551  e/ C y D, Vedado, C. Habana, Cuba.}
\author{Daniel G.\ Barci}
\affiliation{Departamento de F{\'\i}sica Te\'orica,
Universidade do Estado do Rio de Janeiro, Rua S\~ao Francisco Xavier 524, 20550-013,  Rio de Janeiro, RJ, Brazil.}
\date{\today}

\begin{abstract}
We present a functional formalism to derive a generating functional for correlation functions of a multiplicative stochastic process represented by a Langevin equation.
We deduce a path integral over a set of fermionic and bosonic variables without performing any time discretization. The usual prescriptions to define the Wiener integral appear in our formalism in the definition of Green functions in the Grassman sector of the theory. We also study non-perturbative constraints imposed by BRS symmetry and supersymmetry on correlation functions. We show that the specific prescription to define the stochastic process is wholly contained in tadpole diagrams. Therefore, in a supersymmetric theory the stochastic process is uniquely defined since tadpole contributions cancels at all order of perturbation theory.
\end{abstract}

%02.50.-r 	Probability theory, stochastic processes, and statistics (see also section 05 Statistical physics, thermodynamics, and nonlinear dynamical systems)
 
%02.50.Cw 	Probability theory
 
%02.50.Ey 	Stochastic processes
 
%02.50.Fz 	Stochastic analysis
 
%02.50.Ga 	Markov processes 

%05.10.-a 	Computational methods in statistical physics and nonlinear dynamics (see also 02.70.-c in mathematical methods in physics)
 
%05.10.Cc 	Renormalization group methods
 
%05.10.Gg 	Stochastic analysis methods (Fokker-Planck, Langevin, etc.) 

%05.40.-a 	Fluctuation phenomena, random processes, noise, and Brownian motion (for fluctuations in superconductivity, see 74.40.-n; for statistical theory and fluctuations in nuclear reactions, see 24.60.-k; for fluctuations in plasma, see 52.25.Gj; for nonlinear dynamics and chaos, see 05.45.-a)
 
%05.40.Ca 	Noise 

\maketitle

%%%%%%%%%%%%%%%%%%%%%%%%%%%%%%%%%%%%%%%%%%%%%%%%%%%%%%
\section{Introduction}
%%%%%%%%%%%%%%%%%%%%%%%%%%%%%%%%%%%%%%%%%%%%%%%%%%%%%%

Stochastic processes~\cite{Oksendal} has had a great interest in the scientific community since many years ago. Applications can be found along a wide area of scientific research, from physics and chemistry~\cite{vanKampen,Gardiner} through biology and eco\-lo\-gy~\cite{Poschel,Murray} to economy and social sciences~\cite{Mantegna,Bouchaud}.

Perhaps, the most popular formalisms to deal with this type of problems are the Langevin and Fokker-Plank ones~\cite{vanKampen,Gardiner}. While the Langevin description is based on a set of stochastic differential equations, the Fokker-Plank one is based on the probability distribution of the set of variables satisfying a deterministic partial differential equation. The Langevin description rose as a formulation of Brownian motion~\cite{Lubensky-book} and gives a very clear insight into the microscopic properties of the system. In its original application\cite{Brownian-Motion}, the influence of the medium on the diffusive particle is modeled by splitting its effect in a deterministic part, given by the viscous force, and a stochastic part, given by a random force with zero expectation value. In this way, fluctuations exhibit as an additive noise and, consequently, the considered model is an additive stochastic process. In a different case, fluctuations could depend on the state of the system~\cite{Lubensky} and they are regarded as the product of the random force and a function of the state variable. In that situation, multiplicative noise is defined and the model is known as a multiplicative stochastic process.

There is a continuously growing interest in  modeling and studying  noisy systems~\cite{Gitterman} and a great amount of work was developed for additive as well as multiplicative noise. As an example of multiplicative noise, we would like to mention the stochastic Landau-Lifshitz-Gilbert~\cite{Palacios} equation, used to describe dynamics of classical magnetic moments of individual magnetic nanoparticles. Here, the noisy fluctuations of the magnetic field couple the magnetic moment in a multiplicative way. Another interesting example is the problem of state-dependent diffusion~\cite{Lubensky}. The simplest example is the diffusion of a particle in suspension near a wall where the diffusion coefficient depends on the distance to the wall.

We are interested in the path integral formulation of stochastic processes modeled by Langevin equations with multiplicative noise. This is an interesting topic as long as the path integral formalism provides a useful technique to compute correlation and response functions~\cite{Janssen}. Once one is able to built a generating functional for correlation functions, it is possible to study the system using all the machinery developed in quantum field theory, even non-perturbative techniques. In particular, it would be possible to implement a dynamical renormalization group~\cite{Goldenfeld,Janssen-RG} to look for the asymptotic behavior of correlation functions. Within this formalism, symmetries play an essential role since they give information about non-trivial conserved quantities. Moreover, non-trivial symmetries~\cite{SUSy-Work} govern the route to equilibrium of a dynamical system.

The first quantum field theory formalism  describing additive noise was developed by Martin, Siggia and Rose~\cite{MSR} to perturbatively compute  correlation and response functions. After that, the formalism was extensively developed and today, the case of additive noise is reasonably well understood~\cite{Zinn-Justin}. Path integral methods for multiplicative noise have been derived in different situations. 
When the stochastic process is driven by colored noise there is no interpretation problem and a phase-space path integral treatment was developed for Gaussian as well as non-Gaussian shot noise~\cite{Hanggi-shot}. For Gaussian white noise, path integral methods were worked out in the It\^{o}~\cite{Phythian},  Stratonovich~\cite{Graham} and also in more general conventions~\cite{Janssen-RG,Lubensky}.
However, the case of multiplicative noise is more involved and some controversies~\cite{Arnold} have appeared in the literature. Indeed, for fixed stochastic prescription, different results were presented  for the generating functional~\cite{Arnold,Lubensky,Janssen-RG,Zinn-Justin},  depending on the use of continuous or discrete time methods to derive the path integral.

In this paper, we present a derivation of the generating functional using just functional methods, without discretizing the Langevin equation at any intermediate step. We treat additive as well as multiplicative noise in the same footing. The use of auxiliary Grassman variables allows us to describe (at least formally) a stochastic multiplicative process without any reference to the definition of the Wiener integral. In this way, the known prescriptions~\cite{Hanggi} to define the stochastic process, emerge in our formalism in  the definition of Green functions at definite points.  This procedure is useful to study time reversal symmetry and, in general, to study the approach to equilibrium.

In the case of additive processes, there is no ambiguities in the definition of the Wiener integral. However, the explicit construction of a generating functional does depend on the type of discretization adopted~\cite{Hanggi-barrier}. We show that there is no contradiction since, due to a  hidden symmetry (supersymmetry),  the tadpole contributions (carrying the discretization ambiguities) cancel to all order of perturbation theory. 
In addition, supersymmetry enforce time reversal symmetry and fluctuation dissipation relations. This observation may be useful to analyze non trivial multiplicative vector processes.   
We discuss in detail the relation between supersymmetry, equilibrium and the Generalized Stratonovich convention~\cite{Hanggi} (also called  ``$\alpha$-convention''~\cite{Janssen-RG} in the field theory literature) in additive as well as multiplicative stochastic processes.

The paper is organized as follows: In section \ref{Functional} we define our model by means of a Langevin equation with multiplicative noise and derive the generating functional for correlation functions. This is the main result of the paper. In \S \ref{Integration} we show, carefully integrating over Grassman variables, the connection of our result with previous ones obtained by discretization of the Langevin equation. Then, we study the symmetry properties of our action in section \ref{Symmetries}. In particular, we show that in the case of additive noise, where the correlation functions are independent of particular prescriptions,  tadpole diagrams cancel to   all order of perturbation theory due to a hidden supersymmetry. Finally, we discuss our results and future perspectives in section \ref{Discussion}. Appendix \S \ref{AGrassman} briefly summarizes the usual operations with Grassman variables used along the paper.

%%%%%%%%%%%%%%%%%%%%%%%%%%%%%%%%%%%%%%%%%%%%%%%%%%%%%%%%%%%%%%%%%%%%%%%%%%%%%%%%%%%%%%%%%%%%%%
\section{Generating functional}
\label{Functional}
%%%%%%%%%%%%%%%%%%%%%%%%%%%%%%%%%%%%%%%%%%%%%%%%%%%%%%%%%%%%%%%%%%%%%%%%%%%%%%%%%%%%%%%%%%%%%%

For simplicity, we consider a single random variable $x(t)$ satisfying the first order differential equation
\begin{equation}
\dfrac{dx(t)}{dt} = f(x(t)) + g(x(t))\eta(t),
\label{eq.Langevin} 
\end{equation}
where $\eta(t)$ is a Gaussian white noise,
\begin{equation}
 \left\langle \eta(t)\right\rangle   = 0 \;\;\mbox{,}\;\;\;  \left\langle \eta(t)\eta(t')\right\rangle = \delta(t-t').
\label{whitenoise}
\end{equation}
The generalization to a system of variables is straightforward.
In Eq. (\ref{eq.Langevin}),  the drift force $f(x)$ and  the square root of the diffusion function $g(x)$ are,  in principle,  arbitrary smooth functions of $x(t)$. 

To completely define eq. (\ref{eq.Langevin}), it is necessary to give sense to the ill-defined product $g(x(t))\eta(t)$, since $\eta(t)$ is delta correlated. The underline microscopic physics should fix this interpretation using  the It\^o, Stratonovich or even more general prescriptions.  Since our formalism is not based on an explicit discretization of the Langevin equation, we will formally work  independently of any kind of prescription. In fact, the main result, eq. (\ref{action}), does not depend on the explicit definition of the Wiener integral. Of course, the ambiguities will show up in a different way, since they are at the stem of the stochastic process. 
We discuss this problem with great detail in  \S \ref{Integration}.  

We are interested in computing $n-\mbox{points}$ correlation functions $\left\langle x(t_1) \dots x(t_n) \right\rangle$. To compute them, we should know the $n^{\rm th}$-order joint probability  of the random variable $x(t)$. An alternative equivalent procedure is to solve the Langevin equation (Eq. (\ref{eq.Langevin})) and compute, 
\begin{equation}
\left\langle x(t_1) \dots  x(t_n) \right\rangle \equiv  \left\langle \bar x_{[\eta]}(t_1) \dots \bar x_{[\eta]}(t_n) \right\rangle_\eta
\end{equation}
where  $\bar x_{[\eta]}(t)$ is a solution of Eq. (\ref{eq.Langevin}) for a particular realization of the noise and certain initial condition $x(t_0)=x_0$. $<\ldots>_\eta$ means the stochastic mean value in the $\eta$ variable.
If the initial condition is not determine but it is distributed with certain probability ${\cal P}(x_0)$, the expectation value should be taken also in this variable. 

Correlation functions can be obtained from a generating functional 
\begin{equation}
 Z[J(t)] = {\left\langle e^{\int dt J(t) \bar{x}_{[\eta]}(t) }\right\rangle}_\eta
\label{ZJxbar}
\end{equation}
by simply differentiating with respect to the source $J(t)$,
\begin{equation}
 \left\langle \bar x_{[\eta]}(t_1) \dots \bar x_{[\eta]}(t_n) \right\rangle_\eta  = \dfrac{\delta ^n Z[J]}{\delta J(t_n) \dots \delta J(t_1) } \Bigg \vert_{J=0}
\end{equation}
(note that $Z(0)=1$ by definition). 

The aim of this section is to find a functional representation for $Z(J)$ avoiding the problem of explicitly solving the Langevin equation. 
For this purpose we introduce a functional integral over $x(t)$ and a delta-functional, rewriting Eq. (\ref{ZJxbar}) in the following form, 
\begin{equation}
 Z[J(t)] = {\left\langle \int \mathit{Dx(t)}\; \delta[x(t)- \bar{x}_{[\eta]}(t)] \ e^{\int dt J(t) x(t) }\right\rangle }_\eta
%\label{}
\end{equation}
or, since the only noise dependence comes from $\bar x_{[\eta]}(t)$, 
\begin{equation}
 Z[J(t)] =  \int \mathit{Dx(t)} e^{\int dt J(t) x(t) } {\left\langle \delta[x(t)- \bar{x}_{[\eta]}(t)]\right\rangle \ }_\eta .
\label{eq:Zsinruido}
\end{equation}

The following property of the delta-functional is very useful, 
\begin{equation}
\delta[x(t)- \bar{x}_{[\eta]}(t)]=  \delta[\hat{O}(x)]\;\;{\rm det}\left(\frac{\delta \hat{O}}{\delta x}\right)\; ,
\label{delta-property}
\end{equation}
where $\bar{x}_{[\eta]}(t)$ is a root of $\hat{O}(x)$, \emph{i.e.}, $\hat{O}(\bar{x}_{[\eta](t)}) = 0$.

From  Eq. (\ref{eq.Langevin}) we can choose for $\hat{O}(x)$ 
\begin{equation}
\hat{O}(x) = \frac{dx(t)}{dt} - f(x(t)) - g(x(t))\eta (t)
\label{O}
\end{equation}
and correspondingly, the differential operator $\delta \hat{O}/\delta x$ is given by 
\begin{equation}
\dfrac{\delta \hat{O}(x(t))}{\delta x(t')} = \left[ \frac{d}{dt} - f'(x) - g'(x)\eta (t)\right] \delta (t-t'),
\label{O'}
\end{equation}
where $f'$ and $g'$ mean differentiation with respect to $x$. 
Therefore, Eq. (\ref{eq:Zsinruido}) now reads
\begin{eqnarray}
 Z[J(t)]& = & \int \mathit{Dx(t)}\; e^{\int dt J(t) x(t) }\; {\left\langle \delta[\hat O(x)] \;{\rm det}\left(\frac{\delta \hat{O}}{\delta x}\right)\right\rangle \ }_\eta.  \nonumber \\
 & & \label{eq:Z2}
\end{eqnarray}
with the definitions Eqs.\ (\ref{O}) and (\ref{O'}).

It should be noted that for additive noise processes, $g'(x)=0$ and then, the determinant in Eq. (\ref{eq:Z2}) is noise independent.  However, in the general  case, the determinant should be carefully treated.
To do that, we introduce a couple of Grassman functions $\xi(t)$ and $\bar \xi(t)$, obeying the usual Grassman algebra
\begin{equation}
\{\xi(t),\xi(t')\}=\{\bar\xi(t),\bar\xi(t')\}=\{\xi(t),\bar\xi(t')\}=0,
\end{equation}
where $\{~, \}$ represents anticommutators. These relations imply, in particular, $\xi(t)^2=\bar\xi(t)^2=0$.  
The determinant can be represented~\cite{Zinn-Justin} as (for details see Appendix \S \ref{AGrassman}): 
\begin{eqnarray} 
\lefteqn{  {\rm det}\left(\frac{\delta \hat{O}}{\delta x}\right) = \int \mathit{D}\xi \mathit{D}\bar{\xi} \ e^{\int dt dt'\;\bar{\xi}(t) \left( \frac{\delta \hat{O}(x(t))}{\delta x(t')}\right)\xi(t') } }\nonumber  \\
    &=&  \int \mathit{D}\xi \mathit{D}\bar{\xi} \ e^{\int dt \bar{\xi}(t) \frac{d}{dt} \xi(t)} \ e^{-\int dt \bar{\xi}(t)\left[ f'(x) + g'(x)\eta(t)\right] \xi(t)}. \nonumber \\
    & & \label{detgrassman} 
\end{eqnarray}
Note that, different from the additive case, the noise couples to the Grassman variables. 

A last auxiliary function $\varphi(t)$ is introduced to represent the delta-functional

\begin{equation}
\delta\left[\hat O(x) \right] = \int \mathit{D}\varphi \  e^{-i\int dt \  \varphi(t)\left[ \frac{dx(t)}{dt} - f(x) - g(x)\eta (t)\right]}.
\label{delta}
\end{equation}

Substituting the results (\ref{detgrassman}) and (\ref{delta}) into Eq.~(\ref{delta-property}),  we find
%\begin{widetext}
\begin{eqnarray}
\lefteqn{
\left\langle \delta [x(t) - \bar{x}_{[\eta]}(t)]\right\rangle _{\eta}=
\int \mathit{D}\xi \mathit{D}\bar{\xi} \mathit{D}\varphi\;  \times }\nonumber \\
&\times& e^{\int dt\;\left\{ \bar{\xi}(t) \dot\xi(t) -\bar{\xi}(t) \xi(t) f'(x) -i  \varphi(t) \left[\dot x(t) - f(x) \right]\right\}}
\nonumber \\ 
&\times& \left\langle  e^{-\int dt\;\left\{ g'(x) \bar{\xi}(t) \xi(t) -i \varphi(t) g(x) \right\}\eta(t)}\right\rangle _{\eta} \ ,
\label{delta-exvalue}
\end{eqnarray}
%\end{widetext}
where the ``dot'' means time differentiation.
 
With this representation for the delta functional, the average over the noise can now be easily computed. 
We find, 
\begin{equation}
\left\langle  e^{-\int dt\;\left\{ g'\bar{\xi}\xi -i \varphi g\right\}\eta}\right\rangle _{\eta}
=e^{-\int dt\;\left\{ \frac{1}{2}\varphi^2 g^2 + i\varphi g g' \bar{\xi} \xi\right\}}.
\label{average}
\end{equation}
The first term, proportional to $g^2(x)$, is the usual term which comes from the average of the delta-functional. It is also obtained when dealing with an additive process. However, 
the second term, proportional to $g(x) g'(x)$, is proper of multiplicative noise processes and has its origin in the noise dependence of the determinant,  stated in Eq.~(\ref{delta-property}). This term is responsible for the well-known spurious drift terms\cite{Hanggi}. 
One important fact in the current example of a single random variable is the absence of any term proportional to 
$g'(x)^2$. It is automatically canceled due to the property of the Grassman variables, $\xi(t)^2=\bar\xi(t)^2=0$. This situation changes for  a system of stochastic equations with several variables.  

Introducing Eq. (\ref{delta-exvalue}) into Eq. (\ref{eq:Z2}) and using Eq. (\ref{average}) we finally arrive at the desire  representation for the generating functional of correlation functions, 
\begin{equation}
 \mathit{Z}[J] = \int \mathit{Dx}\mathit{D}\varphi\mathit{D}\xi \mathit{D}\bar{\xi} \ e^{-S[x,\varphi,\xi,\bar\xi] + \int dt J(t)x(t)} \ ,
\label{eq:funcionalgenerador}
\end{equation}
where the ``action'' $S$ is given by
\begin{widetext}
\begin{equation}
 S = \int dt \left\lbrace  -\bar{\xi}(t)\frac{d}{dt}\xi(t) + f'(x)\bar{\xi}(t)\xi(t) +\frac{1}{2}\varphi(t)^2 g(x)^2 + i\varphi(t) \left[ \frac{dx}{dt} - f(x) + g(x)g'(x)\bar{\xi}(t)\xi(t)\right]\right\rbrace .
\label{action}
\end{equation}
\end{widetext}

Thus, we have deduced a generating functional for correlation functions as a functional integral over the variables $x(t), \varphi(t), \bar\xi(t), \xi(t)$.
The multiplicative noise effects are evident in the coupling of the Grassman variables with $\varphi(t)$. This coupling is absent in the additive case. 
Usually, the effect of the determinant in Eq.~(\ref{delta-property}) have been treated by discretizing the Langevin equation. In this context, there are some ambiguities in taking the continuum limit that provide, in general, different stochastic evolutions.  In the next section, we will show how these ambiguities appear in our formalism and the relation of our results with previous ones. 

%%%%%%%%%%%%%%%%%%%%%%%%%%%%%%%%%%%%%%%%%%%%%%%%%%%%%%%%%%%%%%%%
\section{Integration over Grassman variables, ambiguities and prescriptions}
\label{Integration}
%%%%%%%%%%%%%%%%%%%%%%%%%%%%%%%%%%%%%%%%%%%%%%%%%%%%%%%%%%%%%%%%

We have obtained the generating functional and its corresponding action, Eqs. (\ref{eq:funcionalgenerador}) and (\ref{action}), by means of functional methods, without any reference to discretization of the Langevin Eq. (\ref{eq.Langevin}).
However, it is well known that different forms of discretization and therefore, different ways to derive the continuum limit, produce different stochastic evolutions. The reason for that behavior resides in the delta-correlated noise. In fact, for white noise, the product $g(x(t))\eta(t)$ (in Eq.\ (\ref{eq.Langevin})) is ill-defined due to the infinite variance of the noise. In many physical applications this problem can be overcame by considering a weakly colored Gaussian-Markov noise with a finite variance\cite{Hanggi-shot}. In this case, there is no problem with the interpretation of eq.~(\ref{eq.Langevin}) and we can take the limit of infinite variance at the end of the calculations. This regularization procedure is equivalent to the so called Stratonovich interpretation~\cite{vanKampen,Zinn-Justin}. However, in other applications, like chemical Langevin equations~\cite{vanKampen} or econometric problems~\cite{Mantegna,Bouchaud},  the noise can be considered principally white, since it could be a reduction of jump-like or Poisson like processes. In such cases, the It\^o interpretation should be more suitable. Therefore, the interpretation of eq.~(\ref{eq.Langevin}) depends on the physics behind a particular application. Once the interpretation is fixed, the stochastic dynamic is unambiguously defined. 

Our formalism is not obviously depending on such interpretations (see eq.~(\ref{action})). In this way, it is specially appropriated to study dynamical properties that are independent of particular prescriptions. In the following,  we show how different interpretation of eq.~(\ref{eq.Langevin}) appear in the Grassman path integral formalism.

The problem can be easily understood looking at the integral
\begin{equation}
\int    g(x(t))\; \eta(t) dt= \int  g(x(t))\;  dW(t) \ ,
\end{equation} 
where we have defined the Wiener process W(t) as $\eta(t)=dW(t)/dt$. 

By definition, the Riemann-Stieltjes integral is
\begin{equation}
\int dt\;  g(x(t))\;  dW(t)=\lim _{n\to\infty} \sum_{j=1}^n  g(x(\tau_j))(W(t_{j+1})-W(t_j))
\end{equation}
where $\tau_j$ is taken in the interval $[t_j,t_{j+1}]$. For a smooth measure $W(t)$, the limit converges to a unique value regardless the value of $\tau_j$.  However, $W(t)$ is not smooth,  in fact, it is nowhere integrable. In any interval, white noise fluctuates an infinite number of times with infinite variance. Therefore, the value of the integral depends on the prescription for the choice of $\tau_j$.
There are several prescriptions to define this integral that can be summarized in the so called Generalized Stratonovich prescription~\cite{Hanggi} or ``$\alpha$-prescription''~\cite{Janssen-RG}, for which we choose $g((1-\alpha)x(t_j)+\alpha x(t_{j+1}))$ with $0\le \alpha \le 1$. In this way, $\alpha=0$ corresponds with the It\^o interpretation and $\alpha=1/2$ coincides with the Stratonovich one.  Moreover, the post-point prescription $\alpha=1$ is also known as the kinetic or the H\"anggi-Klimontovich interpretation~\cite{Hanggi,Hanggi2,Hanggi3,Klimontovich}.

In this section, we show that the same ambiguities appear in the action, Eq. (\ref{action}), without any reference to discretization. The problem is associated with the definition of the determinant of Eq. (\ref{detgrassman}). To compute this formal expression is necessary to define the Green function of the operator $d~/dt$. This problem is common to any integration over Grassman variables and it is not proper of multiplicative noise. As we will see it is also present in any white noise stochastic process. 

For concreteness and to illustrate this point, let us consider the following functional integration over Grassman functions,
\begin{equation}
{\cal I}_G=\int \mathit{D}\bar{\xi} \mathit{D}\xi \ e^{ \int dt\; \bar{\xi}(t) \hat A \xi(t)}\;\; e^{ -\int dt\; \bar{\xi}(t) \xi(t) h(t)} \ ,
\label{GrassmanIntegral}
\end{equation}
where $\hat A$ is any differential operator and $h(t)$ is an arbitrary function of time. 
To compute it, we perform a Taylor expansion of the second exponential, obtaining
\begin{eqnarray}
\lefteqn{
{\cal I}_G=\int \mathit{D}\bar{\xi} \mathit{D}\xi \ e^{ \int dt\; \bar{\xi}(t) \hat A \xi(t)}\; \times } \nonumber \\ 
&\times&\left\{ 1 - \int dt\; \bar{\xi}(t) \xi(t)\; h(t) + \right. \nonumber \\
&+& \left. \frac{1}{2} \int dt dt'\; \bar{\xi}(t) \xi(t) \bar{\xi}(t') \xi(t')\; h(t) h(t') - \ldots \right\}.
\label{Taylor1}
\end{eqnarray}
Computing the Gaussian integral by using Wick theorem we find(see appendix \S \ref{AGrassman})
\begin{eqnarray}
\lefteqn{
{\cal I}_G={\rm Det}\left(\frac{d\;}{dt}\right)\left\{ 1 - A^{-1}(0)\int dt\; h(t) + \right.} \nonumber \\
&+& \left. \frac{1}{2!} \int dt dt' \left[(A^{-1}(0))^2 - A^{-1}(t,t')A^{-1}(t',t) \right] h(t) h(t')\right. \nonumber \\
 &-& \frac{1}{3!} \left. \int dtdt'dt'' \ldots \right\} \ ,
\label{Taylor2}
\end{eqnarray}
where the Green function $A^{-1}(t,t')$ is defined as $\hat A\; A^{-1}(t,t')=\delta(t-t')$.

At this point it is necessary to precisely define the Green function $A^{-1}(t-t')$. We choose a causal prescription by considering the retarded Green function in such a way that
\begin{equation}
A_R^{-1}(t-t') A_R^{-1}(t'-t)=0
\end{equation}
except, possibly, in the null measure set $t=t'$.

To go forward, it is necessary to carefully analyze the function $h(t)$. For delta autocorrelated functions the second line of Eq.~(\ref{Taylor2}) and all the subsequent terms are automatically zero. This is a natural consequence of the anticommuting character of the Grassman variables. In this case, the result is linear in $h(t)$.  However, if  $h(t)$ is a smooth function of time we can safely put $A^{-1}(t-t') A^{-1}(t'-t)=0$ in Eq.~(\ref{Taylor2}) since $t=t'$ is a null measure set. Then, it is  possible to re-exponentiate the Taylor expansion, obtaining 
\begin{equation}
{\cal I}_G={\rm Det}\left(\frac{d\;}{dt}\right)\;\; \exp\{A_R^{-1}(0)\int dt \; h(t)\}.
\label{IG}
\end{equation}

So, the result of Eq.~(\ref{IG}) is valid provided we choose a causal prescription for the Green function of the operator $\hat A$ and we consider smooth functions $h(t)$. Even in this situation, there is an ambiguity, given by $A^{-1}_R(0)$, which has not been defined.  To completely define the determinant we need to fix this value. 

Let us apply this result to integrate the Grassman variables in the generating functional,  Eq.~(\ref{eq:funcionalgenerador}). In this case, the operator $\hat A=\frac{d\;}{dt}$. Therefore, the retarded Green function is $A^{-1}_R=\theta(t-t')$, where $\theta(t)$ is the Heaviside distribution. Therefore, to define the integration we need to fix $\theta(0)=\alpha$. With this choice, the generating functional reads, 

\begin{equation}
\mathit{Z}[J] = \int \mathit{Dx}\mathit{D}\varphi \ e^{-S[x,\varphi] + \int dt J(t)x(t)} \; ,
\label{eq:funcionalgenerador-Lubensky}
\end{equation}
where the ``action'' $S[x,\varphi]$ is given by

\begin{equation}
 S = \int dt \left\lbrace \frac{1}{2}\varphi^2 g^2 + i\varphi \left[ \frac{dx}{dt} - f + \alpha g g'\right]+\alpha f' \right\rbrace.
\label{action-Lubensky}
\end{equation}
Equation~(\ref{action-Lubensky}) coincides with that of  ref.~\cite{Lubensky}, where the parameter $\alpha$ was introduced discretizing the differential equation.  Also,  our action (with $\alpha=1/2$) coincides with the one presented in ref. \cite{Arnold}. However, it differs from that of refs.~\cite{Janssen-RG,Zinn-Justin}. 

To conclude this section, we note that the information about the particular interpretation of eq.~(\ref{eq.Langevin}) is not contained in the action, eq.~(\ref{action}). It is encoded in the definition of the Grassman Green functions at the origin. This is a general result and does not depend on the details of the stochastic differential equation.

We would also like to remark that , even in the case of 
additive noise ($g'=0$), the action does depend on $\alpha$ through the last term in eq. (\ref{action-Lubensky}). However, $Z(J)$ should be $\alpha$ independent ($dZ/d\alpha=0$). Therefore, it should be a non-trivial symmetry of the action  enforcing this property. We address and detail this issue in the next section.  
%%%%%%%%%%%%%%%%%%%%%%%%%%%%%%%%%%%%%%%%%%%%%%%%%%%%%%%
\section{Symmetries and constraints}
%%%%%%%%%%%%%%%%%%%%%%%%%%%%%%%%%%%%%%%%%%%%%%%%%%%%%%%
\label{Symmetries}

The introduction of Grassman variables allows to study symmetries, implemented as linear transformations between the variables $x,\varphi,\xi,\bar\xi$.
These symmetries have direct physical consequences and they strongly constraint the correlation functions, giving information about non-perturbative structure of the theory. Even in the case of perturbation theory, they are useful by helping to rearrange diagrams more efficiently.

%%%%%%%%%%%%%%%%%%%%%%%%%%%%%%%%%%%%%%%%%%%%%%%%%%%%%%%%%%%%%%
\subsection{BRS symmetry}
\label{BRS}

By construction, there is a hidden symmetry in the generating functional, Eq.~(\ref{eq:Z2}), that is related with the fact that $Z(0)=\int {\cal D \eta } \; {\cal P}(\eta)=1$ (see Eq. \ref{ZJxbar})),  \emph{i. e.}, it is a consequence of probability conservation.

From Eq. (\ref{eq:Z2}) it can be readily seen that 
\begin{eqnarray}
Z[0] &=&\left\langle\int \mathit{Dx(t)} \;{\rm det}\left(\frac{\delta \hat{O}}{\delta x}\right)\;\delta[\hat O(x)]\right\rangle_\eta \nonumber \\
&=&\left\langle\int \mathit{D\hat O}\; \delta[\hat O] \right\rangle_\eta =\left\langle 1 \right\rangle_\eta =1 \ , \label{Z0}
\end{eqnarray}
where the measure 
\begin{equation}
\mathit{D\hat O}=\mathit{Dx(t)}\; {\rm det}\left(\frac{\delta \hat{O}}{\delta x}\right) 
\label{measure}
\end{equation}
is trivially invariant under the translation group $\hat O\to \hat O + a$, with constant $a$. 
Since we are working with the  stochastic variable $x(t)$, this symmetry is implemented as $\hat O(x')=\hat O(x)+a$,  that, for infinitesimal transformation, reads 
\begin{equation}
\delta x=x'-x= a \left(\frac{\delta \hat O}{\delta x}\right)^{-1},
\label{nlt}
\end{equation}
where $a$ is an infinitesimal parameter and $\left(\frac{\delta \hat O}{\delta x}\right)^{-1}$ is the inverse operator of Eq.~(\ref{O'}). 
Thus, the invariance of the measure, Eq.~(\ref{measure}), under the highly non-linear transformation $\delta x$, guarantees the probability conservation property $Z(0)=1$. In this way, it is clear the role played by the determinant as a Jacobian of the non-linear change of variables $\hat O=\hat O(x)$. 

The representation of the determinant in terms of Grassman variables allows to  visualize the non-linear transformation, Eq. (\ref{nlt}), as a linear transformation between variables in the extended space $(x,\varphi, \xi, \bar\xi)$.
Indeed, it can be shown that, even after the integration over the noise,  the generating functional, Eq.~({\ref{eq:funcionalgenerador}}), is invariant under the linear transformations
\begin{eqnarray}
\delta x &=& \bar\lambda \xi\mbox{~~~~~,~~~~~~} \delta \xi=0 \ ,  \label{BRS1}\\
\delta \bar\xi &=& i\bar\lambda \varphi\mbox{~~~~,~~~~~~} \delta \varphi=0 \ , \label{BRS2} 
\end{eqnarray}
where $\bar\lambda$ is an anticommuting parameter.  This nilpotent transformation  ($\delta^2=0$) is the famous BRS~\cite{BRS} symmetry, discovered in the context of quantization of Gauge theories.

A more convenient representation is obtained by introducing a new Grassman (``temporal'') coordinate $\bar \theta$ and the functions 
\begin{eqnarray}
X(t,\bar\theta)&=& x(t)+  \bar\theta \xi(t), \label{X} \\
\bar C(t,\bar\theta)&=&\bar\xi(t) + i\bar\theta \varphi(t).
\label{C}
\end{eqnarray}
In terms of these functions, the transformations in Eqs. (\ref{BRS1}) and (\ref{BRS2}) are simply translations in  $\bar\theta$. In fact,
\begin{equation}
\delta X(\bar\theta)=\bar\lambda \;{\cal D}X(\bar\theta)\mbox{~~~,~~~}\delta \bar C(\bar\theta)=\bar\lambda\; {\cal D}\bar C(\bar\theta) \ ,
\end{equation}
where the generator ${\cal D}=\partial /\partial \bar\theta$, satisfying ${\cal D}^2=0$.

The action  of Eq.~(\ref{action}) can be written in terms of $X$ and $\bar C$ in such a way that the BRS symmetry displays in an obvious way. 
We can write the action, before integrating over the noise, in the following way: 
\begin{equation}
S=\int dt d\bar\theta\; \bar C(\bar\theta)\left(\dot X(\bar\theta) - f(X(\bar\theta)) - g(X(\bar\theta))\;\eta(t)  \right).
\end{equation}
It is interesting to note that, in the generating functional, the formal functional integration over  $\bar C$ enforce the equation 
\begin{equation}
\dot X(\bar\theta) - f(X(\bar\theta)) - g(X(\bar\theta))\;\eta(t) = 0  ,
\end{equation} 
which has the same (formal) structure of the Langevin equation, Eq.~(\ref{eq.Langevin}). 

Integrating over the noise, we find
\begin{eqnarray}
S & = & \int dt d\bar\theta\; \bar C(\bar\theta)\left(\dot X(\bar\theta) - f(X(\bar\theta))\right) \nonumber \\
& + & \int dt d\bar\theta_1 d\bar\theta_2\; \bar C(\bar\theta_1)\bar C(\bar\theta_2)g(X(\bar\theta_1))g(X(\bar\theta_2)).
\label{action2}
\end{eqnarray}
The advantage of this expression is that it is explicitly BRS invariant, since  its dependence on $\bar\theta$ is only implicit thorough the functions $X$ and $\bar C$. Of course, integrating Eq.~(\ref{action2}) over $d\bar\theta$, we find Eq.~(\ref{action}) as we should.

From the representation of the determinant, Eq.~(\ref{detgrassman}), we observe that the Grassman variables always appear in pairs $\bar\xi\xi$. Assigning a ``fermionic number'' $+1$ to the variable $\bar \xi$ and $-1$ to $\xi$,  each term of any correlation function must have zero fermionic number, {\em i. e.}, there is fermionic number conservation. This can be seen in  eq.~(\ref{action2}) by the fact that for each $d\bar\theta$ there is one function $\bar C$. Fermionic number conservation imposes 
\begin{equation}
\langle \xi(t_1)x(t_2) \rangle=\langle \xi(t_1)\varphi(t_2) \rangle=\langle \xi(t_1)\xi(t_2) \rangle=0 ,  
\label{fnc}
\end{equation}
therefore, $\langle X(t_1,\bar\theta_1)X(t_2,\bar\theta_2)\rangle=\langle x(t_1)x(t_2)\rangle$.

On the other hand, since the system is invariant under translations in $\bar\theta$, any correlation function should be a function of $\bar\theta_2-\bar\theta_1$. For instance, the correlation $\langle X\bar C\rangle$ takes the general form
\begin{equation}
\langle X(t_1,\bar\theta_1)\bar C(t_2,\bar\theta_2) \rangle= A(t_1,t_2) + \left(\bar\theta_2-\bar\theta_1\right) B(t_1,t_2) ,
\label{XC}
\end{equation} 
where $A$ and $B$ are functions of time.
Combining this  equation with the definitions of $X$ and $\bar C$, Eqs.~(\ref{X})-(\ref{C}),  and considering the fermionic number conservation, Eq.~(\ref{fnc}), we find 
\begin{equation}
\langle x(t_1)\varphi(t_2)\rangle=\langle \bar\xi(t_1)\xi(t_2)\rangle,
\label{constraint}
\end{equation}
in such a way that 
\begin{equation}
\langle X(t_1,\bar\theta_1)\bar C(t_2,\bar\theta_2) \rangle=  \left(\bar\theta_2-\bar\theta_1\right) \langle x(t_1)\varphi(t_2)\rangle.
\label{XC1}
\end{equation} 
Eq.~(\ref{constraint}) is an important constraint imposed by BRS symmetry and it is valid to all order of perturbation theory. It states that the correlation functions of Grassman variables are equal to the physical response of the system to a delta perturbation.

We have seen in the last section that, in order to integrate over Grassman variables, it is necessary to define the Green function. We chose the retarded prescription, supplemented with $A^{-1}_R(t,t)=\alpha$. Eq.~(\ref{constraint}) imposes the same behavior for the response functions, in fact, $\langle x(t_1)\varphi(t_2)\rangle=0$ for $t_1<t_2$  and $\langle x(t)\varphi(t)\rangle=\alpha$.
Thus, we are not free to choose any other prescription for the responses. If, for some reason, we decide to use another convention we would be breaking BRS symmetry and, in that way, we would be spoiling the probability conservation.

%%%%%%%%%%%%%%%%%%%%%%%%%%%%%%%%%%%%%%%%%%%%%%%%%%%%%%%%%%%%%
\subsection{Supersymmetry and tadpole cancellation}
\label{Supersymmetry}

There is another convenient representation of the same action that reveals new symmetries and constraints. Let us introduce a new Grassman variable $\theta$, conjugate to the previous one $\bar\theta$ and define the superfield
\begin{equation}
\Phi(t,\theta,\bar\theta) = X(t,\bar\theta) + \bar C(t, \bar\theta) \theta.
\end{equation}
It is simple to see that
\begin{eqnarray}
\bar C(\bar\theta) & = & - \frac{\partial \Phi}{\partial \theta} \mbox~{~~~,~~~~}
f(X(\bar\theta))= \int d\theta\; \theta f(\Phi).
\end{eqnarray}
Then, Eq.~(\ref{action2}) can be rewritten as a local action in the superspace $\{t,\theta,\bar\theta\}$
\begin{equation}
S=\int dtd^2\theta\left[\theta \frac{\partial \Phi}{\partial\theta}\left\{\frac{\partial\Phi}{\partial t} - f(\Phi)\right\} + \frac{1}{2} \frac{\partial\Phi}{\partial\theta}\frac{\partial\Phi}{\partial\bar\theta} g^2(\Phi)\right]
\label{action-multiplicative}
\end{equation}
In this representation, BRS symmetry is again evident since the action does not depend explicitly on $\bar\theta$ (note the asymmetric role of $\bar\theta$ and $\theta$).
The correlation function $\langle\Phi(\theta_1,\bar\theta_1,t_1)\Phi(\theta_2,\bar\theta_2,t_2)\rangle$,  which contains  the physical correlation functions between $x(t),\varphi(t),\xi(t),\bar\xi(t)$, is strongly constrained. Due to the BRS symmetry,  causality and fermionic number conservation,  a general two point correlation function, has the form
\begin{eqnarray}
\lefteqn{
\langle\Phi_1\Phi_2\rangle= \langle x(t_1)x(t_2)\rangle+ (\bar\theta_1-\bar\theta_2) \times} \nonumber \\
&\times &
\left[(\theta_1-\theta_2) \langle x(t_1)\varphi(t_2)\rangle_- +(\theta_1+\theta_2) \langle x(t_1)\varphi(t_2)\rangle_+ \right],\nonumber \\
& &
\label{structure}
\end{eqnarray}
where $\mp$ represents retarded and advanced prescriptions, respectively. 

Taking into account that the parameter $\alpha=\langle  x(t)\varphi(t)\rangle=\langle \bar\xi(t) \xi(t)\rangle$, any correlation function depends on $\alpha$ only through tadpole diagrams in a perturbative expansion. In the superspace formulation, tadpole diagrams represent products of superfields and their derivatives at the same superspace point $(t,\theta,\bar\theta)$. 

Interestingly, from Eq.~(\ref{structure}) it is immediate to check the property~\cite{Olemskoi} 
\begin{equation}
\int dt d\theta d\bar \theta\;\; \langle \Phi(t,\theta,\bar\theta)\Phi(t,\theta,\bar\theta)\rangle=0 \ ,
\label{tadpole}
\end{equation}
which has deep consequences in the dynamics of the system.  To see this point, let us consider the example of a Langevin equation with additive noise, \emph{i.e.}, $g(\Phi)=g=\mbox{constant}$. In this situation, there is another BRS symmetry related with the variable $\theta$, which constraints even more the correlation functions. In fact, it is possible to rewrite the action, Eq.~(\ref{action-multiplicative}), to make this symmetry explicit. In the case of one random variable, it is always possible to write $f(\Phi)=\frac{\partial V(\Phi)}{\partial\Phi}$. 
Then, 
\begin{equation}
S=\int dt d\theta d\bar\theta\; \left\{\bar D\Phi D\Phi +V(\Phi)\right\},
\label{action-additive}
\end{equation}
where we have defined 
\begin{equation}
\bar D=\frac{\partial~}{\partial\theta} \mbox{~~~,~~~} D = \frac{1}{2} g^2 \frac{\partial~}{\partial\bar\theta}-\theta\frac{\partial~}{\partial t} \ ,
\end{equation}
which satisfy  $D^2=\bar D^2=0$ and $\left\{D,\bar D\right\}=-\frac{\partial~}{\partial t}$.

The action of Eq. (\ref{action-additive}) has an extended symmetry (called supersymmetry (SUSY)~\cite{SUSY}), whose generators are
\begin{equation}
 D'= \frac{\partial~}{\partial\bar\theta} \mbox{~~,~~} \bar D'= \frac{1}{2}g^2\frac{\partial~}{\partial\theta}+\bar\theta\frac{\partial~}{\partial t}
\mbox{~~,~~} \left\{D',\bar D' \right\}=\frac{\partial~}{\partial t} \ .
\end{equation}
The first generator, $D'$ coincides with the previous BRS one. In addition, there are two more generators associated with translations in time and in $\theta$.
The graded algebra 
\begin{eqnarray}
\left\{D', D \right\}&=&\left\{D', \bar D \right\}=\left\{\bar D', D \right\}=\left\{\bar D', \bar D \right\}=0 , \\
D'^2 &=& \bar D'^2=0 \ ,
\end{eqnarray}
guarantees that the variation of the action is a total derivative. Supersymmetry has very important physical as well as mathematical implications. From the physical point of view, it ensures that the stochastic process reaches the thermodynamical equilibrium at long times~\cite{Zinn-Justin, SUSy-Work}. 
In fact, SUSY implies that two-point correlation functions depend on the time difference. On the other hand, it enforces the fluctuation dissipation theorem~\cite{Leticia}.

From the mathematical point of view, noting  that $V(\Phi)$ does not depend on ``velocities''  (nor time, neither $\theta$ derivatives) and using the general equation Eq.~(\ref{tadpole}), it is immediate to conclude that the contribution of tadpole super-diagrams cancels identically to all order of perturbation theory.  
This is a general result in any supersymmetric theory~\cite{SUSY}. It is based on the fact that the Hamiltonian ({\em i. e.}, the generator of time translations, $H$) can be cast as an anticommutator of SUSY generators and,  since the ground state is supersymmetric, $\langle H \rangle=0$.

As long as there are no more diagrams, other than tadpole ones, which contain $\langle  x(t)\varphi(t)\rangle=\langle \bar\xi(t) \xi(t)\rangle=\alpha$, we conclude that correlations do not depend on the prescription to define Green functions of differential operators. SUSY guarantees that the stochastic dynamic is unique and tends to equilibrium configurations at long times, independently of any prescription to define the Wiener integral. Therefore, correlation functions are $\alpha$ independent at any order of perturbation theory. 

In the multiplicative noise case, we have no such a symmetry for general functions $f(x)$ and $g(x)$ (Eq.~(\ref{action-multiplicative})). 
Although Eq. (\ref{tadpole}) is valid, because it only depends on BRS symmetry, there will appear tadpole diagrams with derivative couplings like 
$\langle \frac{\partial \Phi}{\partial\theta }\frac{\partial \Phi}{\partial\bar\theta }\rangle$ that do not cancel and therefore,  correlation functions will depend on $\alpha$. In this situation, each prescription $\alpha$ defines a different stochastic process. Thus, the coupling of the Grassman variables with the stochastic variable $x(t)$ through $g g' \bar\xi \xi$ gives rise to the usual spurious drift terms. It should be noted that, in the case of several variables (vector variables), the spurious drift term could be zero for a special class of diffusion matrices~\cite{Hanggi}. In this case, the correlation functions do not depend on $\alpha$ even for the case of multiplicative noise. Whether this behavior implies a supersymmetry in the path integral approach is an interesting open question.

%%%%%%%%%%%%%%%%%%%%%%%%%%%%%%%%%%%%%%%%%%%%%%%%%%%%%%%%%%%%%%%%%%
\section{Discussion and conclusions}
\label{Discussion}
%%%%%%%%%%%%%%%%%%%%%%%%%%%%%%%%%%%%%%%%%%%%%%%%%%%%%%%%%%%%%%%%%%

We have developed a functional formalism to describe stochastic processes modeled by Langevin equations with multiplicative noise. We have treated additive as well as multiplicative noise at the same level. Our  derivation of the generating functional for correlation and response functions does not rely on the discretization of  the Langevin equation at any intermediate step. In this way, we did not  need to explicitly face the ambiguities inherent to the definition of a Wiener integral. Of course, since these ambiguities are at the stem of the stochastic process, they reappear in our formalism in a different context, \emph{i.e.}, in the precise definition of Green functions in the Grassman sector. 

We found a generating functional written in terms of a functional integral over bosonic as well as fermionic (Grassman) variables (Eqs. (\ref{eq:funcionalgenerador}) and (\ref{action})). We have also shown that, carefully integrating over the Grassman variables, we reproduce previous results~\cite{Lubensky,Arnold} obtained by discretizing the Langevin equation (Eq.\ (\ref{action-Lubensky})).  

In particular, we have shown that the results achieved by using the ``$\alpha$-convention''  are reobtained in our formalism by just fixing the retarded  Green function for Grassman variables, $G_R(t,t)=\alpha$. In our example, the differential operator is $\frac{d~}{dt}$ and the retarded Green function is the Heaviside distribution $\theta(t-t')$. However, the method is general and can be applied to any differential operator. 

One of the practical advantages of this formalism is that it is not necessary to take care of the  effects that prescriptions make on different stochastic evolution. It is possible to  work with the same mathematical structure and, at the end of the calculation, the equal time Green function is fixed, whether it is necessary. 

Also, the presence of Grassman variables makes the symmetries of the theory evident. 
When the action is formulated in superspace, in terms of $\Phi(t,\theta,\bar\theta)$, it is clear that the stochastic prescriptions are related with  tadpole super-diagrams.  Moreover, if the theory is supersymmetric, the tadpole diagrams cancel identically to all order of perturbation theory. Therefore, in the special case of a supersymmetric theory, all prescriptions to define the Wiener integral conduce to the same stochastic evolution. On the other hand, from a physical point of view, supersymmetry implies that the stochastic evolution tends asymptotically to the thermodynamic  equilibrium~\cite{Gozzi}.

We believe that our approach clarifies previous controversies~\cite{Arnold} about the difference between discrete and continuum approaches to the path integral formulation of multiplicative noise.  On the other hand, it provides a natural unified framework to deal with  different prescriptions used to define stochastic processes.
The implementation of BRS symmetry and supersymmetry in a linear form allows to compute restrictions to correlation functions in a non-perturbative way, helping to guide actual computations in concrete models.

The formalism presented in this paper can be readily generalized to more general cases as, for instance, to vector stochastic variables. It could also be possible to treat systems driven by white shot noise.  We hope to report on those issues briefly. 
 
\acknowledgments
The Brazilian agencies {\em Conselho Nacional de Desenvolvimento Cient\'{\i}fico e Tecnol\'{o}gico (CNPq)},
and the {\em Funda{\c {c}}{\~{a}}o de Amparo {\`{a}} Pesquisa do Estado do Rio de Janeiro (FAPERJ)}  are acknowledged for partial  financial support.
This work was supported by the Latin American Centre of Physics, CLAF, under the collaborative program CLAF-ICTP, and by  the  Academy of Sciences for the Developing World, TWAS.

\appendix
\section{Representation of a determinant in terms of Grassman integration}
\label{AGrassman}

We briefly review in this appendix some usual manipulations with Grassman variables that we have used along the paper.  
For a detailed treatment of this subject in statistical mechanics as well as in quantum field theory we refer the reader to ref. \onlinecite{Zinn-Justin}. 

We define a set of $n$ Grassman variables $\theta_i$ and its conjugates $\bar\theta_i$, where $i=1\ldots n$ as 
\begin{equation}
\{\theta_i,\theta_j\}=\{\bar\theta_i,\bar\theta_j\}=\{\theta_i,\bar\theta_j\}=0 
\end{equation}  
where $\{a,b\}=ab+ba$ is the anticommutator of $a$ and $b$. This definition implies the nilpotent property $\theta_i^2=\bar\theta_i^2=0$. Therefore, any function of these variables should be a polynomial of degree at least $2n$. For instance, if $n=1$, 
\begin{equation}
F(\theta,\bar\theta)=A + B\;\theta + C\;\bar\theta + D\;\theta\bar\theta
\label{Fn1}
\end{equation}  
is the most general function of $\theta$ and $\bar\theta$. The coefficients $A,B,C,D$ are  complex numbers. 

Differentiation in the Grassman variable is also a nilpotent operator $\partial^2/\partial^2\theta=0$ satisfying the Clifford algebra
\begin{equation}
\{\frac{\partial~}{\partial\theta_i},\frac{\partial~}{\partial\theta_j}\}=0,\;\;\; 
\{\frac{\partial~}{\partial\theta_i},\theta_j\}=\delta_{ij} 
\end{equation}  
and its conjugates. Any anticommutator that mixes $\theta_i$ and $\bar\theta_j$ is also zero since they are independent. 

Interestingly, integration in a Grassman space is the same operation as differentiation. Therefore, the use of a derivative or an integral symbol is a matter of taste. For instance, taking into account eq. (\ref{Fn1}),  
\begin{equation}
\int d\theta d\bar\theta\; F(\theta, \bar\theta)=D
\label{integral}
\end{equation}
That means that the integration over $d\theta d\bar\theta$ picks up the coefficient of the term $\theta\bar\theta$. 

Let us consider the Gaussian Grassman integral 
\begin{equation}
I_G(A)=\int \left(\prod_{k=1}^{n}d\theta_k d\bar\theta_k\right)\;\;
e^{\sum_{ij}\bar\theta_i A_{ij} \theta_j }
\label{IGA}
\end{equation}
where $A_{ij}$ are the elements of a matrix $A$.
According with eq. (\ref{integral}), the integral is the coefficient of the term proportional to $\theta_1\bar\theta_1\theta_2\bar\theta_2\ldots\theta_n\bar\theta_n$. 
Therefore, expanding the exponential in eq. (\ref{IGA}) in a ``finite'' Taylor series, and reordering the terms taking into account the anticommuting properties of the Grassman variables, 
\begin{equation}
I_G(A)=\sum_{j_1,j_2,\ldots, j_n} (-1)^{P} A_{n,j_n}A_{n-1,j_{n-1}}\ldots A_{1,j_1}
\end{equation}
where $P$ is the order of the permutation of $\{j_1,\ldots, j_n\}$. We immediately recognize that 
\begin{equation}
I_G(A)=\det(A)
\label{detA}
\end{equation}
This result should be compare with the output of a normal Gaussian integral that is $I(A)\sim {\det^{-1}(A)}$. Therefore, 
in the same way that the inverse of an  $n\times n$ matrix determinant  can be represented as a Gaussian integral over a set of $n$ complex variables, the determinant itself can be represented as a Gaussian integral over a  set of $n$ complex Grassman variables.

We can generalize now the case of a discrete set of Grassman variables $\{\theta_1,\ldots\theta_n\}$, to an infinite set of continuous variables; {\it i.\ e.\ }, a Grassman function $\xi(t)$.  In this case we can generalize eqs. (\ref{IGA}) and (\ref{detA}) to
\begin{equation}
\det(A)=\int \mathit{D}\xi \mathit{D}\bar{\xi} \ e^{\int dt dt'\;\bar{\xi}(t) A(t,t')\xi(t')}
\end{equation} 
where $\det(A)$ is a functional determinant and $\mathit{D}\xi \mathit{D}\bar{\xi}$ are functional integrations over Grassman variables.
$A(t,t')$ is the kernel of the functional $A$. 
This is the formula used in eq. (\ref{detgrassman}) to represent the Jacobian $\delta\hat O/\delta x$.

\end{document}